\documentclass{article}

\usepackage{graphicx}

\usepackage{lineno,hyperref}
\usepackage{url}
\usepackage{breakurl}
\usepackage[utf8]{inputenc}
\usepackage{amsmath}
\usepackage{latexsym}
\usepackage[english]{babel}
\usepackage{amsfonts}
\usepackage{subfig}
\usepackage{mathptmx}

\newtheorem{definition}{Definition}

\begin{document}

\title{A feasibility study for a persistent homology based $\mathbf{k}$-Nearest Neighbor search algorithm in melanoma detection}

\author{Massimo Ferri\\
	Universit\`{a} di Bologna - Bologna, Italy, \textsf{massimo.ferri@unibo.it} \and
        Ivan Tomba\\
        CA-MI S.r.l., Pilastro - Parma, Italy, \textsf{tomba.ivan@gmail.com} \and
        Andrea Visotti\\
        CA-MI S.r.l., Pilastro - Parma, Italy, \textsf{andrea.visotti@gmail.com} \and
        Ignazio Stanganelli\\ 
        IRCCS-IRST, Meldola - Forlì, Italy and \\
        Università di Parma - Parma, Italy, \textsf{igstanga@tin.it}
}

\maketitle

\begin{abstract}
Persistent Homology is a fairly new branch of Computational Topology which combines geometry and to\-pology for an effective shape description of use in Pattern Recognition. 
In particular it registers through ``Betti Numbers'' the presence of holes and their persistence while a parameter (``filtering function'') is varied.
In this paper, some recent developments in this field are integrated in a $k$-Nearest Neighbor search algorithm suited for an automatic retrieval of melanocytic lesions.
Since long, dermatologists use five morphological parameters (A $=$ Asymmetry, B $=$ Boundary, C $=$ Color, D $=$ Diameter, E $=$ Elevation or Evolution) for assessing the malignancy of a lesion. 
The algorithm is based on a qualitative assessment of the segmented images by computing both 1 and 2-dimensional Persistent Betti Numbers functions related to the ABCDE
parameters and to the internal texture of the lesion.
The results of a feasibility test on a set of $107$ melanocytic lesions are reported in the section dedicated to the numerical experiments.
\end{abstract}
 
\section*{Biographies}

\textbf{Massimo Ferri} is full professor of Geometry at the University of Bologna. 
Coming from topology of low-dimensional manifolds, he grew interested in the applications of geometry and topology to robotics and to pattern recognition. His present research is in persistent topology and applications.

\noindent \textbf{Ivan Tomba} received his Ph.D. degree in Mathematics from the University of Bologna in 2013. He is currently software developer for CA-MI S.r.l. and inventor (with Andrea
Visotti) of a patent pending video-dermatoscope with a software for an automatic image retrieval of melanocytic lesions. His research interests include Inverse Problems and 
geometrical-topological methods for shape comparison and applications in Pattern Recognition.

\noindent \textbf{Andrea Visotti} has been the R$\&$D Coordinator of CA-MI S.r.l. for two years, during which he filed
two patent applications and a model utility as inventor. 
He is also inventor (with Ivan Tomba)
of a patent pending video-dermatoscope with a software for an automatic image retrieval of mela\-no\-cy\-tic
lesions. Currently he is consultant
for CA-MI S.r.l. and co-founder and R$\&$D Coordinator of IBD S.r.l. to reverse engineer and innovate the total spectrum of biomedical machines with most of the functionalities
but at a much lower cost.

\noindent \textbf{Ignazio Stanganelli} is director of the Skin Cancer Unit IRCCS Istituto Scientifico Romagnolo per la cura e lo studio dei tumori (IRST) and associate professor of Dermatology
Department at University of Parma. His research is mainly focused on the development of new non-invasive technologies such as digital dermoscopy, computer aided diagnosis and confocal
laser microscopy in the skin tumors diagnosis, especially in the secondary prevention of melanoma.

\section{Introduction}
\label{intro}
The incidence of malignant melanoma in fair-skinned patients has increased dramatically in most parts of the world over the past few decades.\\
According to the World Health Organization ($2006$), $132$ thousands melanomas occur globally each year and more than $65$ thousands people a year worldwide die for melano\-ma \cite{Lucas06}.\\
According to the American Cancer Society, $6.640$ males and $3.300$ 
females are expected to die of melanoma in $2015$ and the incidence of melanoma has increased $15$ times in the last $40$ years \cite{Wein2001}. In the United Kingdom and Europe, a similar increase in the incidence has been seen. In the European Union, the incidence of invasive melanoma is of $9$ cases over $100.000$ every year \cite{Ferlay08}. 
This incidence tends to increase with the latitude, with a larger prevalence among the populations with a lower pigmentation (from $12$ to $17$ cases over $100.000$ every year) than 
among the Mediterranean countries (from $3$ to $5$ cases over $100.000$ every year).
Moreover, melanoma is the second most frequent neoplasm among people under 
$40$ years old, underlining that the problem is often related to the patients' young age \cite{Siegel14}.\\
Since the prognosis of melanoma depends almost entirely on tumor thickness, detection of early melanoma is crucial for the survival of patients \cite{Kopf94,Stanga01}.
The widely used acronym ABCDE (asymmetry, irregular borders, multiple colors, diameter $>$ 6 mm, enlarging lesion) contains the primary clinical criteria for diagnosing 
suspected skin melanoma. \\
However, the early phase of malignant melanoma is difficult to identify by naked eye because cutaneous malignant melanoma can share many clinical features
with an atypical naevus. Several studies have described diagnostic accuracy rates ranging from $50$ to $75\%$, 
indicating a need for additional diagnostic tools \cite{DelMar94,Miller92,Masood13}. \\
Dermoscopy (or epiluminescence microscopy) is the examination of skin lesions with a dermatoscope. This traditionally consists of a magnifier, a non-polarised light source,
a transparent plate and a liquid medium between the instrument and the skin. It is a noninvasive method that allows the in vivo evaluation of colors and microstructures of the
epidermis, the dermoepidermal junction and the papillary dermis not visible to the naked eye. These structures are specifically correlated to histologic features. 
The identification of specific diagnostic patterns related to the distribution of colors and dermoscopy structures can better suggest a malignant or benign pigmented skin lesion. 
The use of this technique provides a valuable aid in diagnosing pigmented skin lesions \cite{Massone05,Pehamberger87,Pehamberger93}.\\
Because of the complexity involved, this methodology is reserved for experienced clinicians \cite{Stanga98,Stanga99,Stanga98Bis}. In \cite{Vestergaard}
is reported that dermoscopy assessment is more accurate than clinical evaluation by naked eye for the diagnosis of cutaneous melanoma ($OR = 15.6$, $p = 0.016$). 
In this study, the mean sensitivity in the diagnosis of melanoma was $74\%$ for the examination by naked eye and $90\%$ for dermoscopy.\\
The acquisition of dermoscopic images has also stimulated the automatization of the diagnostic process: many rather successful computer programs have been implemen\-ted 
to automatically analyze melanocytic lesions, providing the dermatologist with a support in the diagnosis phase \cite{Burroni04,D'Amico04,Masood13,Seiden99,Stanga05}.
These programs are generally based on a software that given an image of a skin lesion from a test set (query image) and a training set of already classified images performs the
following four steps on the query:
\begin{itemize}
 \item[1)] removal of artifacts (hair, bubbles, ...);
 \item[2)] segmentation: isolation of the skin lesion from its background;
 \item[3)] analysis: feature extraction;
 \item[4)] classification: assignment of the query to one of the clas\-ses in which the training set is divided.  
\end{itemize}
Many different techniques such as decision trees, support vector machine, extreme learning machine, statistical methods, the $k$-nearest neighbor ($k$-NN) and many others
have been developed to perform the classification step, with a very wide range of results in terms of sensitivity, specificity and diagnostic accuracy:
see \cite{Masood13} for a summary of the results obtained by many different algorithms in the last two decades; see also \cite{Dreis09} and \cite{Fried09} for a comparison
between automatic and human performance.\\
In the analysis step, most of the programs keep into account the traditional ABCDE parameters used by dermatologists: Asymmetry (of boundary, texture, and color), Boundary 
(irregularity and dishomogeneity), Color (presence of several colors), Dimension, and Evolution. An original method for comparing in a qualitative, yet precise way
two skin lesions is based on the mathematical theory of Persistent Homology (originated as Size Theory). 
This method was introduced in the early 90's to compare homeomorphic topological spaces and has considerably grown in popularity since it has been
proven to provide both theoretical and computational tools for shape comparison \cite{Biasotti11,Biasotti08}. The main idea is to take
into account topological shape features with respect to some geometric properties conveyed by real functions defined on the shape itself. Typically, a shape
is represented by a pair $(\mathcal{X}, \varphi)$, where $\mathcal{X}$ is a topological space and $\varphi : \mathcal{X} \rightarrow \mathbb{R}$ is a continuous real-valued function
called filtering (or measuring) function. Persistent Homology allows to associate to the pair $(\mathcal{X}, \varphi)$ some shape descriptors (Persistent Betti Number Functions)
which register quantitatively the behaviour of the filtering function. This allows to compare pairs of the type $(\mathcal{X}, \varphi)$ and 
$(\mathcal{Y}, \psi)$, with $\mathcal{X}$ and $\mathcal{Y}$ homeomorphic, by computing a distance between their Persistent Betti Number (PBN) functions.
The Research Group of Vision Mathematics of the University of Bologna has obtained interesting results using these techniques, as proved in a series of papers \cite{D'Amico04,FerriStanga2010,Stanga05}.\\
Recent developments in the field of Persistent Homology led the authors to the implementation of a $k$-NN algorithm that integrates some recent discoveries for the retrieval of skin
lesions.\\
The choice of a $k$-NN algorithm permits retrieval and visualization of the ``most similar'' cases to those at hand. This aspect
partly resembles the medical reasoning and allows a dermatologist to directly compare unknown lesions with other known skin lesions. This can provide an advantage in areas
where black-box models are inadequate, which is exactly the case of many borderline skin lesions. Moreover, retrieval provides a valid support in the diagnosis decision, without
necessarily producing an automatic classification, which tends to influence the user's judgement. On the other hand, as stated in \cite{Masood13}, the major difficulty of 
$k$-NN algorithms lies in the definition of a metric that measures the distance between data items. Another problem is that whereas classification can focus on the search for some
characteristic features, retrieval needs a formalization of the concept of similarity which has to be much more adherent to the intuitive concept with this name: a poor performance
in retrieval is as risky as in classification, but much more evident just by sight.\\
The authors believe that Persistent Homology may be an adequate technical tool in the transition from a risk-assessment output based on classification to image retrieval. However, since
this transition is delicate, as a first step the new algorithm has been evaluated as a classifier. Thus, the aim of this work is to:
\begin{itemize}
 \item describe the new algorithm;
 \item present the numerical results obtained in a preliminary feasibility test made on 107 dermoscopic melanocytic lesions;
 \item evaluate which of the new technical tools may be worth of further investigation.
\end{itemize}   
The paper is organized as follows: section $\ref{PreProcSec}$ describes the pre-processing steps (hair removal and segmentation); section $\ref{analysis}$ presents the analysis step; section 
$\ref{Classification}$ discusses the problems of the choice of the metrics and classification; section $\ref{NumSec}$ is dedicated to the numerical experiments and their discussion
and section $\ref{Conclusions}$ to the conclusions.

\section{Pre-processing}\label{PreProcSec}
\begin{figure}[t]
 \centering
  \subfloat[][Original image.]
{\includegraphics[width=0.45\columnwidth]{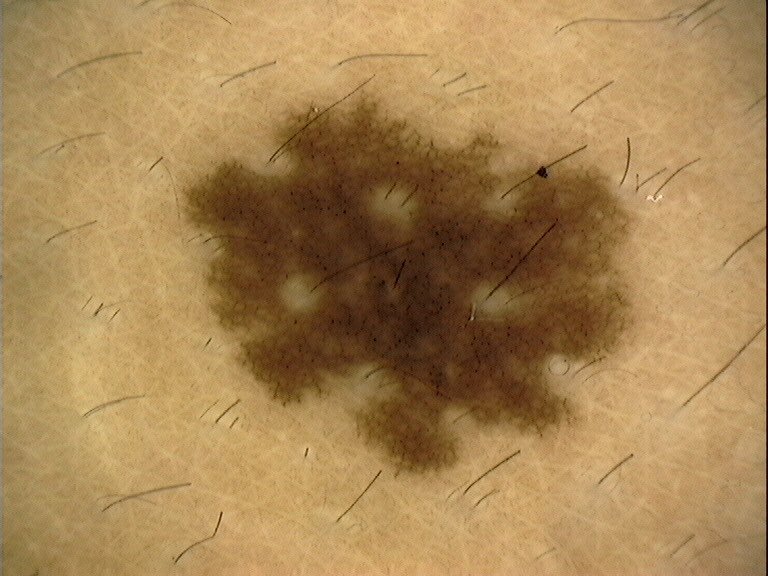}}
 \subfloat[][Output of the hair-removal algorithm.]
{\includegraphics[width=0.45\columnwidth]{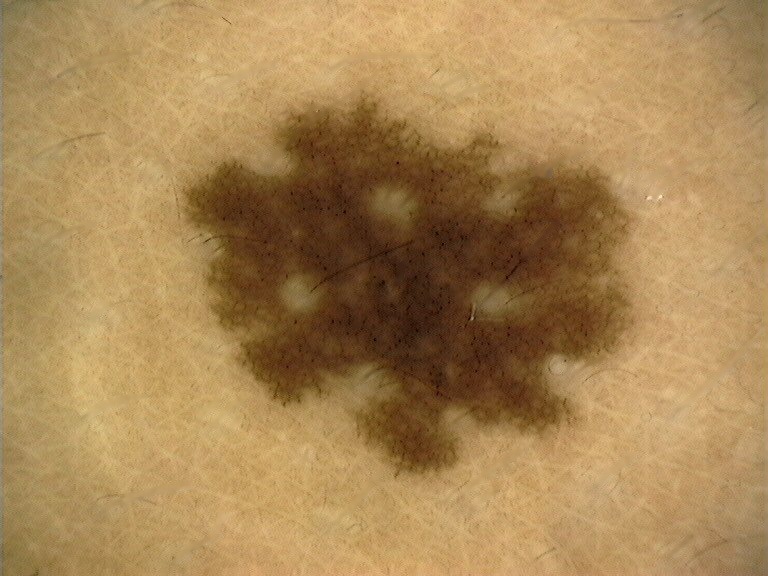}}
 \caption{Example of the hair-removal algorithm.}
 \label{Depilatore}
\end{figure}

\begin{figure}[t]
 \centering
\includegraphics[width=228pt,height=156pt]{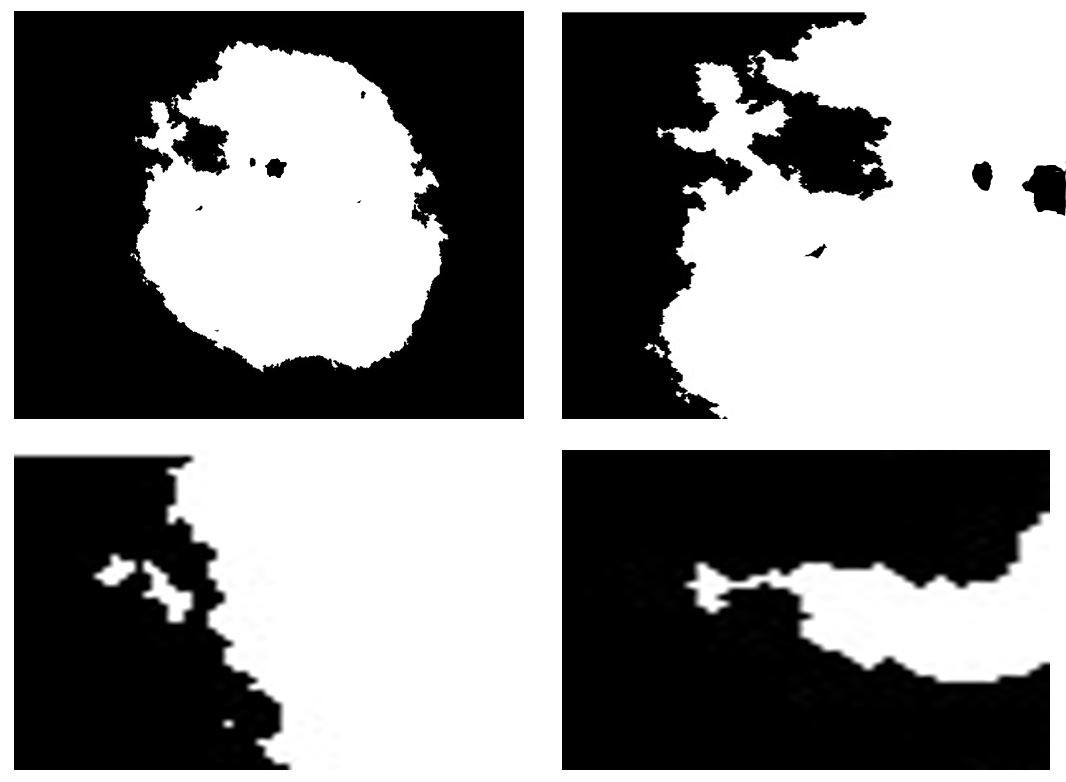}
 \caption{An output mask of the DRLSE algorithm (top left). The top-right figure enlights some holes in the main connected component of the mask;
 the bottom-left figure shows some smaller connected components of the mask; the bottom-right figure shows that the contour curve is not simple,
 due to the presence of an isthmus of boundary pixels which connects two distinct connected components of the inside.}
 \label{Maschere}
\end{figure}

\begin{figure}[ht]
\centering
  \subfloat[][Pre-processing output.] 
{\includegraphics[width=0.45\columnwidth]{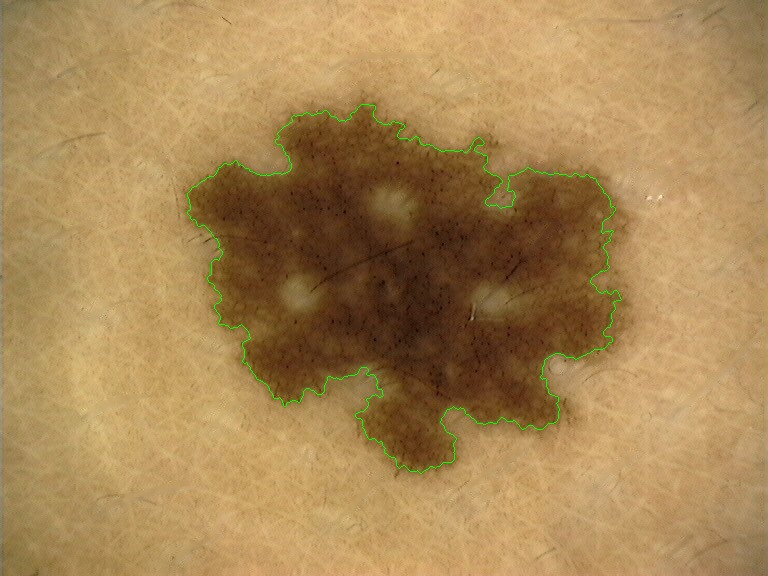}}
 \subfloat[][Its mask.]
{\includegraphics[width=0.45\columnwidth]{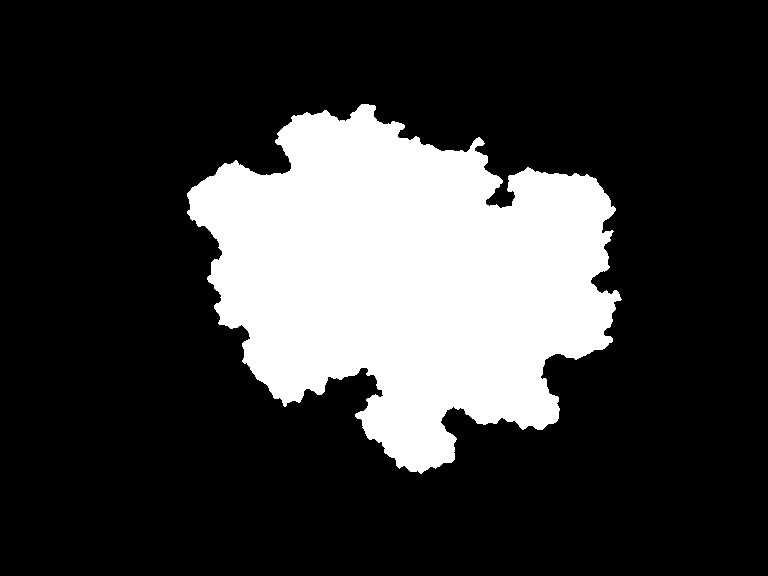}}
 \caption{The final result of the pre-processing algorithm for the image of Figure $\ref{Depilatore}$ and the corresponding mask, which satisfies properties $P1$-$P4$.}
 \label{PreProcFig}
\end{figure}

The pre-processing phase consists of three distinct steps.\\
As a first pre-processing optional step, the software performs the removal of the hair present on the lesion.
The areas that contain the hair are identified by means of an erosion/dilation with straight-line segments in a similar way to that proposed in \cite{Lee97}. This process
terminates with the creation of a Boolean mask containing the hair zones and some noise. For removing the noise, the mask is then treated in a similar way to that described
in \cite{Kiani11}, obtaining a new Boolean mask. At last, the pixels inside this mask are replaced with the average over the neighboring pixels not belonging to the mask.
Figure $\ref{Depilatore}$ shows the result obtained in a practical example, where the picture on the left is the original image acquired by the dermatoscope and
the picture on the left is the output of the hair removal algorithm.\\
In the second step, the aim of the algorithm is to identify the discrete (digital) versions of the two topological spaces that will be associated to
the skin lesions: the boundary of the lesion and the lesion itself. This is achieved by a segmentation algorithm. The segmentation determines the edge of the lesion, identified
by a closed curve that separates the lesion’s area from the background. It is performed as follows.\\
The image is cropped by the user, for framing the skin lesion in an optimal way. Then, a level set evolution type algorithm is applied.\\
Level set methods were introduced by Osher and Sethian in \cite{OshSet88} to study front propagations. Standard formulations consider a region $\Omega \subset \mathbb{R}^2$, a
closed, continuous curve $\pmb{\Gamma}(-,t):$ $[0,1]$ $\rightarrow$ $\Omega$ evolving in time $t$ $\in$ $[0,+\infty)$ and a real-valued function $\Phi:$ $\Omega$ $\times$ 
$[0,+\infty)$, usually called the level set function, such that $\Phi(\pmb{\Gamma}(q,t),t)=0$ for every $q$ $\in$ $[0,1]$ and for every $t \geq 0$.
In these formulations, the level set function typically develops irregularities during its evolution, which may cause numerical errors and eventually destroy
the stability of the evolution \cite{Sethian99}. Therefore, a numerical remedy, called reinitialization, is typically applied to periodically replace the degraded level set function
with a signed distance function \cite{Sethian99}. However, the practice of reinitialization affects numerical accuracy and efficiency in an undesirable way \cite{Sethian99,Gomes00}.
For this reason, in a series of papers \cite{Li08,Li10,Li11}, C. Li et al. proposed a new variational level set formulation (known as DRLSE - Distance Regularized Level Set Evolution) in which
the regularity of the level set function is intrinsically maintained during the level set evolution. The level set evolution is derived as the gradient flow that minimizes
an energy functional with a distance regularization term and an external energy that drives the motion of the zero level set toward desired locations. The distance regularization
effect eliminates the need for reinitialization and thereby avoids its induced numerical errors. This led the authors to the implementation of the DRLSE method described in \cite{Li11},
with satisfying results, in the sense that the algorithm computed the segmentation in real-time and the outcome was in line with the clinicians' expectations for the images considered
in the numerical experiments.\\
The output of the DRLSE algorithm is a Boolean mask containing all those pixels on which the level set function computed by the DRLSE is negative. This mask will represent the
digital space associated to the skin lesion and the border of the mask will be the digital space associated to its border. \\
In general, in skin lesion images, the DRLSE algorithm produces masks constituted by a large connected component containing the most of the skin lesion
and by several very small connected components near the boundary of the main component: a typical example is shown in Figure $\ref{Maschere}$. 
Unfortu\-nately, dealing with objects with a different number of connected components does not
fit one of the main assumptions one usually makes when comparing two sets using persistent homology, i.e. that the underlying topological spaces are homeomorphic 
(cf. \cite{Biasotti08,Biasotti11}).\\
In the continuous $2$-dimensional model, the simplest way to deal with homeomorphic spaces with homeomorphic boundaries is to consider the standard situation in which the sets to
be compared are compact topological spaces whose boundary is a Jordan curve in $\mathbb{R}^2$. The Jordan Curve Theorem ensures that the spaces are homeomorphic. \\
In the discrete setting, the analogous of the Jordan Curve Theorem is a classic result of digital topology usually called the Discrete Jordan Curve Theorem \cite{Eckh94,Kong89}. The
main point behind the Discrete Curve Theorem is to use different connectivity notions for the boundary and the inside, in order to avoid the so-called connectivity paradoxa 
\cite{Rosen79}. The digital version of the continuous properties of the Jordan Curve Theorem can be summarized as follows.
\begin{definition}
 Let $I$ and $J$ be respectively the number of rows and columns of an image $\mathsf{I}$. A function $$\mathsf{M}: \{0,...,I-1\} \times \{0,...,J-1\} 
 \longrightarrow \{-1,0,1\}$$ is said to be a mask of $\mathsf{I}$ suitable for comparisons if it satisfies the following properties:
 \begin{enumerate}
  \item[$P1$] $\mathsf{M}$ is surjective. In this case, a pixel $(i,j)$ will be said to belong to the background if $\mathsf{M}(i,j)=1$, to the inside if 
  $\mathsf{M}(i,j)=-1$ and to the boundary if $\mathsf{M}(i,j)=0$. 
 \item[$P2$] The background is $4$-connected and contains the border of the window (i.e. all the pixels such that $i=0$, $i=I-1$, $j=0$, or $j=J-1$).
 \item[$P3$] The inside is $4$-connected and is $4$-disconnected from the background.
 \item[$P4$] The boundary is $8$-connected and if a pixel belongs to the boundary, it has two and only two $8$-neighbors in the boundary. 
 \end{enumerate}
 \end{definition}
In general, the output mask of the DRLSE algorithm is not suitable for comparisons (cf. Figure $\ref{Maschere}$).\\
This motivates the introduction of a third step in the pre-processing phase, which transforms the DRLSE output mask into a suitable for comparisons mask as follows.
\begin{subsection}{Cleanmask algorithm}
Let $$\mathsf{M}_0: \{0,...,I-1\} \times \{0,...,J-1\} \longrightarrow \{-1,1\}$$ be the mask obtained as the output of the DRLSE algorithm. The Cleanmask
Algorithm will be defined as the consecutive execution of the following actions:
\begin{enumerate}
 \item[\textbf{A1}] copy the mask $\mathsf{M}_0$ on a mask $\mathsf{M}$ setting to the background all the pixels at the border of the window;
 \item[\textbf{A2}] identify a maximal connected component (in the sense of $4$-connectivity) of $\mathsf{M}^{-1}(\{-1\})$ by means of a Depth First Search (DFS) 
 and set every pixel not belonging to this maximal connected component to the background ($\mathsf{M}(i,j)=1$);
 \item[\textbf{A3}] identify the connected components (in the sense of $4$-connectivity) of the resulting background by means of a DFS and in particular the component containing the
 boundary of the image. Set all the pixels not belonging to this component to the inside ($\mathsf{M}(i,j)=-1$);
 \item[\textbf{A4}] for every pixel $(i,j)$ s.t. $\mathsf{M}(i,j)=-1$, if $(i,j)$ has at least a $4$-neighbor belonging to the background, set $\mathsf{M}(i,j)=0$;
 \item[\textbf{A5}] identify a maximal connected component of $\mathsf{M}^{-1}(\{-1\})$ and set all the pixels of $\mathsf{M}^{-1}(\{-1\})$ belonging to
 the other components to the background;
 \item[\textbf{A6}] for every $(i,j)$ s.t. $\mathsf{M}(i,j)=0$, if $(i,j)$ has no neighbors belonging to the inside, set $\mathsf{M}(i,j)=1$.
\end{enumerate}
The output of the Cleanmask Algorithm satisfies the properties $P1$-$P4$ thanks to the Discrete Jordan Curve Theorem.\\
\end{subsection}
The connectivity of the mask and of the boundary (achieved by actions $\mathbf{A2}$ and $\mathbf{A3}$ respectively) are essential assumptions of the persistent homology based
algorithm described in the following sections, since they ensure that the distances between pairs of segmented skin lesions remain finite (see also section \ref{analysis} and 
\cite{Biasotti08,Biasotti11}).\\
In addiction, filling the holes of the mask (action $\mathbf{A3}$) is important from the clinical point of view: it allows to restore those parts of the skin lesion that 
were eliminated by the DRLSE algorithm because the skin was not dark enough there, but that might be of some significance for the diagnosis. For example, a light bluish mark on a 
dark pigmented skin lesion may be a sign of melanoma.\\
Actions $\mathbf{A5}$ and $\mathbf{A6}$ ensure that the boundary is a simple, closed curve in the digital plane. These operations are not actually necessary for the 
persistent-homology algorithm to be well defined, since only $0$-homology will be considered. Moreover, the fact that some parts of the skin lesion segmented by the DRLSE algorithm
may be deleted due to these actions is a possible cause of concern (for example, in the bottom-right picture of Figure $\ref{Maschere}$ the western region of that 
portion of the mask will be eliminated). However, according to the experiments, in practice only very small parts of the lesion are deleted in this phase.
On the other hand, this provides contour curves that are topologically equivalent in the discrete setting (indeed, they are Jordan curves in the digital plane in the sense of 
\cite{Khal90}), resembling the continuous setting in a better way and allowing potential extensions of the comparisons to homology groups of degree $>0$.\\
Summing up, the software performs hair-removal and computes the segmentation of the skin lesion by means of the DRLSE algorithm corrected by the Cleanmask Algorithm, to obtain a mask
which is suitable for comparisons. The final result of the pre-processing phase for the image of Figure $\ref{Depilatore}$ is shown in Figure $\ref{PreProcFig}$, which shows the
segmented image (left picture) and the corresponding suitable for comparisons mask (right picture). The green line in the left picture shows the simple closed boundary contour of
the mask.

\section{Analysis}\label{analysis}
\begin{figure}[t]
 \centering
 \includegraphics[width=228pt,height=156pt]{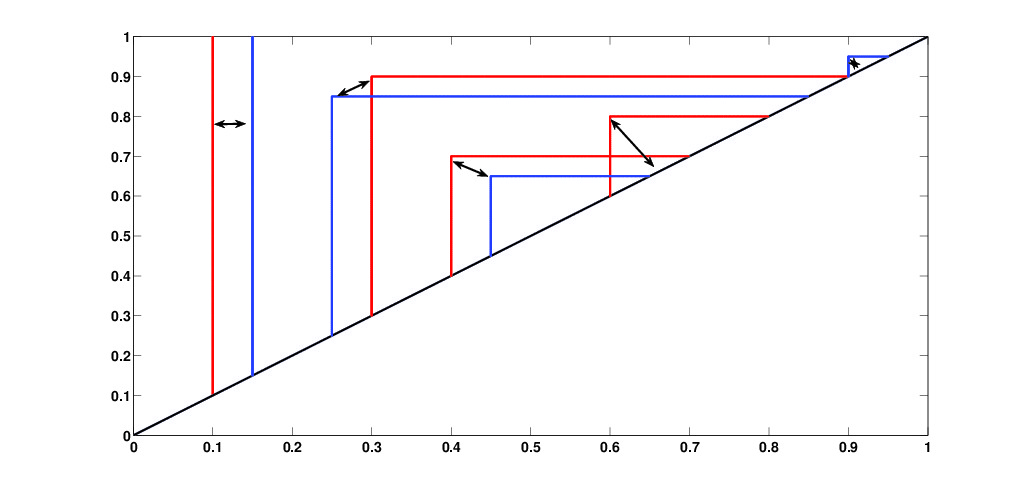}
 \caption{The matching distance between two PBNs in the half-plane is obtained by searching the best matching between the two sets of cornerpoints.}
 \label{Bottle}
\end{figure}

To the aim of comparing images, the main tool the algorithm relies on is the theory of Size Functions (SFs) \cite{Fro99}, extended in Persistent Homology \cite{Cohen05,Edel08}
as Persistent Betti Number functions, also in the form of Persistence Diagrams (PDs).\\
Essentially, the PBN registers the behavior of the filtering function by using Morse theory \cite{Fro96}.
The main idea is to take into account topological shape features with respect to some geometric properties conveyed by real functions defined on the shape itself. 
In formal settings, that means a shape is represented by a size pair $(\mathcal{X},\phi)$, where $\mathcal{X}$ is a topological space and $\phi: \mathcal{X} \longrightarrow \mathbb{R}$
is a continuous real-valued function called filtering (or measuring) function. \\
Size Theory was introduced in the early 90’s to store quantitatively some qualitative information about shapes. In particular, the ($1$-dimensional) 0-PBN (or SF) of the size pair 
$(\mathcal{X},\phi)$ is the function $\ell_{(\mathcal{X},\phi)}$ defined on the half-plane $$\Delta^+:=\{(u,v) \in \mathbb{R}^2 \text{ }|\text{ } v>u \}$$ s.t. for every point
$(u,v)$ $\in$ $\Delta^+$ $\ell_{(\mathcal{X},\phi)} (u,v)$ is the number of path-connected components of the sub-level set
$$
 \mathcal{M}_v:=\{x \in \mathcal{X} \text{ }\|\text{ } \phi(x) \leq v\}
$$
which contain at least one point of the corresponding sub-level set $\mathcal{M}_u$.\\
In Persistent Homology theory, $k$-PBNs with generic integers $k$ $\geq 0$ have also been studied in detail (see e.g. \cite{Edel08} for a formal definition of the $k$-PBNs and
\cite{Cagl10} for a simple example in which $1$-PBNs play an essential role). However, since in practical applications $k$-PBNs with $k>0$ are rarely used as they are not easy to
calculate, in this paper only $0$-PBNs (later, for simplicity, PBNs) will be considered.\\
\begin{figure}[t]
 \centering
{\includegraphics[width=0.45\columnwidth]{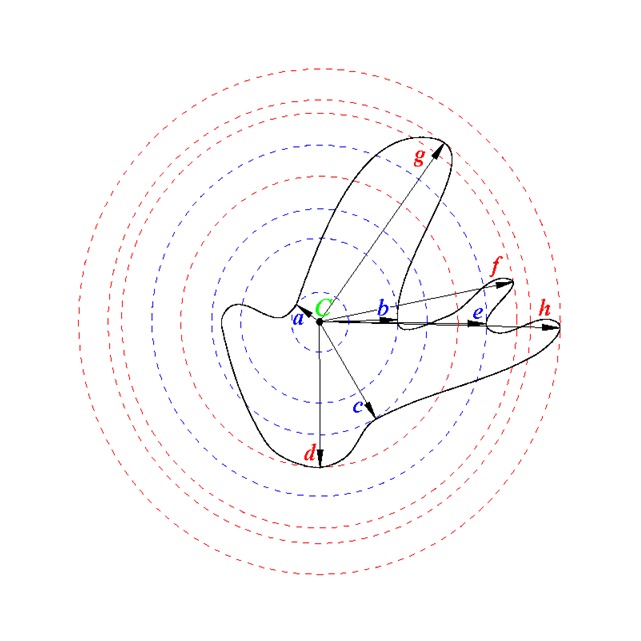}}
{\includegraphics[width=0.45\columnwidth]{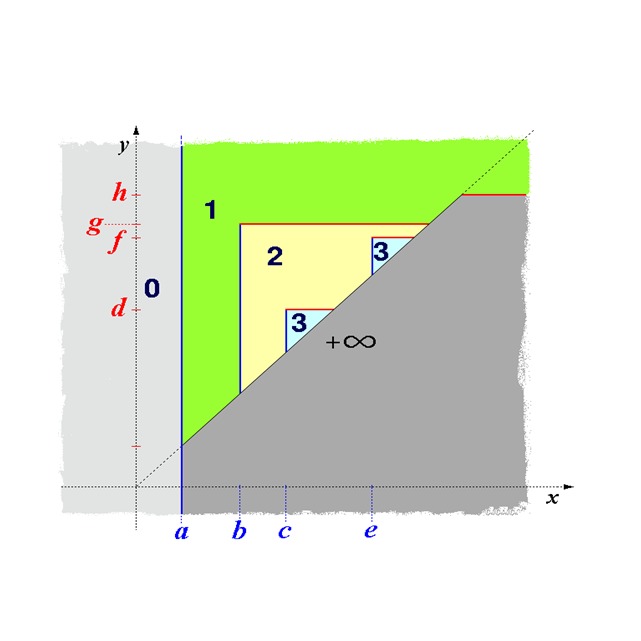}}
 \caption{The computation of the $0$-PBN function of the size pair represented by a Jordan Curve (black line in the left picture) with the Euclidean distance from the point $C$ as
 filtering function. The dotted circles show the radial values that influence the generation of the cornerpoints of the PBN function (right picture).}
 \label{BoundarySF}
\end{figure}
Figure $\ref{BoundarySF}$ shows the computation of a PBN function: in this situation, $\mathcal{X}$ is the (black) Jordan curve in the left picture and the function $\phi$ is the
Euclidean distance from the point $C$ (center of the dotted circles). The picture on the right represents the PBN of this size pair: the numbers on the various coloured areas
of $\Delta^+$ indicate the values that the PBN function assumes on these zones. \\
In general, a PBN function can be seen as a linear combination (with natural numbers as coefficients) of characteristic functions associated to the (possibly unbounded) triangles
laying on $\Delta^+$ (cf. \cite{Fro01}). The bounded triangles are of the form $\{(u, v ) \in \Delta^+ : \alpha \leq u < v < \beta \}$, while the unbounded ones are of the form
$\{(u, v ) \in \Delta^+ : \eta \leq u < v \}$. Hence, a simple and compact representation is obtained by associating the set $\{(u, v ) \in \Delta^+ : \alpha \leq u < v < \beta\}$ 
to the point $(\alpha, \beta)$, and the set $\{(u, v ) \in \Delta^+ : \eta \leq u < v \}$ to the point at infinity $(\eta, \infty)$. The points of a formal series having 
finite coordinates are called proper cornerpoints, while the ones with a coordinate at infinity are said to be cornerpoints at infinity or cornerlines.\\
Persistent Betti Number functions can be compared by using a suitable matching (also called bottleneck) distance
\cite{D'Amico06,D'Amico10}, and have been widely applied to Pattern Recognition and dermoscopy-related problems (see 
\cite{Biasotti08,Cerri06,Dibos04,D'Amico04,FerriStanga2010,Stanga05}).
The matching distance $d_{match}$ can be seen as a measure of the cost of transporting the cornerpoints of a PBN function into the cornerpoints of another one. Formally, let $\ell_1$
and $\ell_2$ be two PBN functions, and let $C_1$ and $C_2$ be the multisets of their cornerpoints counted with their multiplicities and augmented
by adding the points of the diagonal $\Delta:=\{(u,v) \in  \mathbb{R}^2: \text{ } u=v\}$ counted with infinite multiplicity. Denoting by $\overline{\Delta^*}$ the set 
$\overline{\Delta^+}$ extended by the points at infinity of the kind $(a,\infty)$, $a$ $\in$ $\mathbb{R}$, i.e. $\overline{\Delta^*}:=\overline{\Delta^+} \cup \{(a,\infty),
a \in \mathbb{R} \}$, the matching distance is defined as
\begin{equation}
 d_{match}(\ell_1,\ell_2):=\min_{\sigma} \max_{P \in C_1} \delta(P,\sigma(P)),
\end{equation}
where $\sigma$ varies among all the bijections from $C_1$ to $C_2$ and the distance $\delta((u,v),(u'v'))$ between two points $(u,v)$ and $(u',v')$ in $\overline{\Delta^*}$ is
defined by
$$\min \{ \max \{|u-u'|,|v-v'|\}, \max \{\frac{v-u}{2},\frac{v'-u'}{2} \} \},$$
with the convention about $\infty$ that $\infty-v$ $=$ $v-\infty$ $=\infty$ when $v \neq \infty$, $\infty-\infty$ $=$ $0$, $\infty/2$ $=$ $\infty$, $|\infty|$ $=$ $\infty$, 
$\min \{c, \infty\}$ $=$ $c$, $\max \{c, \infty\}$ $=$ $\infty$.
A consequence of this convention is that if $\ell_1$ and $\ell_2$ have a different number of cornerlines, their matching distance is automatically $\infty$. Indeed,
it is easy to see that the number of cornerlines is equal to the number of connected components of the space $\mathcal{X}$, meaning that the distance between two spaces with a 
different number of connected components would be equal to $\infty$.\\
The pseudometric $\delta$ measures the pseudo-distance between two points $(u,v)$ and $(u',v')$ as the minimum between the cost of moving
one point onto the other and the cost of moving both points onto the diagonal, with respect to the max-norm and under the assumption that any two points of the diagonal have
vanishing pseudo-distance.\\
As different PBN functions may in general have a different number of cornerpoints, $d_{match}$ allows a proper cornerpoint to
be matched to a point of the diagonal: this matching can be interpreted as the deletion of a proper cornerpoint.\\
Figure $\ref{Bottle}$ visualizes the computation of a matching distance between two very simple PBN functions, both constituted by a cornerline and three cornerpoints 
(the first PBN is represented in red, the second in blue): after matching the cornerlines, the minimum cost is achieved by matching together two couples of points that are
sufficiently close to each other and by matching the remaining cornerpoints to the diagonal (i.e. deleting them).\\
The success of the matching distance is mainly due to its stability properties. In order to understand these properties, it is common to introduce the natural pseudo-distance
between two size pairs $(\mathcal{X}, \phi)$ and $(\mathcal{Y}, \psi)$, with $\mathcal{X}$ and  $\mathcal{Y}$ homeomorphic, as the quantity:
\begin{equation}
 d((\mathcal{X}, \phi),(\mathcal{Y}, \psi)):=\min_{f} \max_{x \in \mathcal{X}} |\phi(x)-\psi(f(x))|, 
\end{equation}
where $f$ varies among all possible homeomorphisms between $\mathcal{X}$ and $\mathcal{Y}$. The natural pseudo-distance is a measure of the dissimilarity between the two size pairs,
but it is computable only in a very few cases. It turns out that the matching distance is stable with respect to perturbations of the measuring function in the following sense:
\begin{equation}
 d_{match}(\ell_1,\ell_2) \leq d((\mathcal{X}, \phi),(\mathcal{Y}, \psi)).
\end{equation}
Moreover, it has been proved that $d_{match}$ provides the best lower bound for the natural pseudo-distance, in the sense that any other distance between size functions would furnish
a worse bound. For details about these results, the interested reader is addressed to \cite{Biasotti08,Biasotti11,D'Amico06,D'Amico10} and the references therein.\\
In the last decade, the theory of PBNs has developed extensively: in particular, a multi-dimensional PBNs theory is now available. In this context, filtering functions with values in
$\mathbb{R}^k$ ($k \geq 1$) defined on the space $\mathcal{X}$ generate $k$-dimensional PBN functions and the generalization of the domain $\Delta^+$
lies in $\mathbb{R}^{2k}$. The multidimensional matching distance has been proved to improve the lower bounds achieved by the $1$-di\-mensional PBNs separately 
(Proposition $4$ in \cite{Biasotti08}) and in the case $k=2$ a stable algorithm to compute the $2$-dimensional matching distance between $2$-dimensional PBNs is now a\-vailable
\cite{Biasotti11,Cagl10}.\\
To the authors' knowledge, no attempts to apply these results in the dermatology field have been made before this work. Here, the computation of $1$ and $2$-dimensional PBNs given
an $RGB$ skin lesion's image is performed following the approach described below.\\
Since the algorithm deals with digital images, it is important to discuss Persistent Homology and Size Theory in the digital world. The discrete version of the
theory substitutes the topological space $\mathcal{X}$ with a graph $G=(V,E)$, the function $\phi: \mathcal{X} \longrightarrow \mathbb{R}$ with a function $\phi: V \longrightarrow
\mathbb{R}$ and the concept of topological connectedness with the usual connectedness notion for graphs \cite{D'Amico00}. Indeed, the size pair ($\mathcal{X},\phi$) can be obtained
from the discrete data by taking a geometric realization $|V|$ of the graph that encodes the non-zero elements of the image as nodes and the neighborhood adjacency among the digital
points as edges. Depending on the number of neighbors that may be adjacent to a point, the connectivity (i.e. the number of edges) of the graph depends on the number of neighbors
that are admitted to be adjacent to a point in the 2D image (i.e., $8$-, $6$- or $4$-neighborhoods). Then, the function $\phi$ : $|V|$ $\rightarrow$ $\mathbb{R}$
is a piecewise linear function first defined on the nodes of the graph and then linearly extended to the edges. The pair ($G,\phi$) is usually called the Size Graph.
The PBN functions are obtained from the Size Graphs by means of the algorithm described in \cite{D'Amico00}. This algorithm directly computes the multiset of
cornerpoints and cornerlines that completely determines a PBN function in $O(n \log n + m \alpha(2m + n, n))$ operations, where $n$ and $m$ are the number of vertices and
edges in the Size Graph, respectively, and $\alpha$ is the inverse of the Ackermann function \cite{Acker28}.\\
The matching distance in a single half-plane is computed by means of a Hopcroft-Karp type algorithm, whose complexity is $O(p^{2.5})$, where $p$ is the number of cornerpoints taken
into account for the comparison.\\
The evaluation of the $2$-dimensional matching distance between $2$-dimensional PBN functions requires the computation of infinitely many $1$-dimensional matching distances, one for each
point of a certain rectangle of $\mathbb{R}^2$, whose dimensions depend on the filtering function \cite{Biasotti11,Biasotti08,Cagl10}. Since in practice it is impossible to consider 
every point of this rectangle, the algorithm proposed in $\cite{Biasotti11}$ computes a stable approximation of this $2$-dimensional matching distance on a grid defined on the 
rectangle: the denser the grid, the smaller the error. Indeed, the bound for this error is an input of this algorithm: in order to evaluate the performances of the $2$-dimensional
algorithm in this dermatology application, in the experiments considered in this paper this bound has always been chosen small enough so that this error could be considered 
negligible.\\
Focusing on the case of digital images of pigmented skin lesions, the mask obtained by the DRLSE algorithm is turned into a suitable for comparisons mask as described in section 
$\ref{PreProcSec}$. Then, $2$ different graphs are computed: one (hereafter called the Boundary Graph) contains only the boundary pixels and the other one (hereafter
called the Global Graph) contains all pixels of the mask. The connectivity notion on these graphs is given by the $8$-connectivity for the case of the Boundary Graph and by
the $6$-connectivity for the case of the Global Graph. It should be noted that, in line with the continuous setting case, the number of cornerlines computed by the algorithm in
\cite{D'Amico00} is equal to the number of connected components of the Size Graph. This explains the necessity to consider Size Graphs with only one connected component,
in order to avoid cases in which the matching distance is equal to $\infty$: actions $\mathbf{A2}$ and $\mathbf{A3}$ in the Cleanmask Algorithm of section $\ref{PreProcSec}$
are necessary to achieve this goal.\\
In order to compute PBN functions, the following filtering functions have been defined:
\begin{itemize}
 \item[1] on the Boundary Graph, the Euclidean distance from the barycenter of the mask;
 \item[2] on the Global Graph: 
 \begin{itemize}
 \item  the blue (B), greeen (G) and red (R) channels, defined on each pixel (node of the graph)
 as the values (in the range $[0,255]$) assumed by the $RGB$ image on that pixel; 
 \item the light intensity variations, which for a pixel $(i,j)$ is defined as
 $$TV((i,j)):=\sum_{(i',j')} |In((i,j))-In((i',j'))|,$$
 where $In=(R+G+B)/3$ is the light intensity and $(i',j')$ varies in the set of pixels of the mask $8$-con\-nected to $(i,j)$;
 \item  the excess-colours $ExcB$, $ExcG$,
 $ExcR$, defined by formulas of the type
 $$ExcB:= (2B-G-R+510)/4;$$
 \item the three colour-differences $R-G$, $R-B$ and $G-B$, defined by formulas of the type
$$R-B:=(R-B+255)/2.$$
 \end{itemize}
 
\end{itemize}
This provides $11$ different $1$-dimensional Size Graphs $(G,\phi)$ to compute the corresponding $1$-dimensional PBNs.\\
To the aim of computing $2$-dimensional matching distances, $2$-dimensional Size Graphs have been formed by coupling some $1$-dimensional filtering functions. In particular,
the following $2$-dimensional filtering functions have been defined on the Global Graph: 
\begin{itemize}
 \item colour channel couplings:\\ $(B,G)$, $(B,R)$, $(G,R)$;
 \item colour channel and opposite colour difference couplings:\\ $(B,R-G)$, $(G,R-B)$, $(R,G-B)$;
 \item excess-colours and opposite colour difference couplings:\\ $(ExcB,R-G)$, $(ExcG,R-B)$, $(ExcR,G-B)$.
\end{itemize}
Since the Global Graph is usually too large, in order to reduce the computational costs, prior to these operations, the images' resolution has been reduced by splitting them into
small squared blocks and by taking averages of the three color channels on these blocks. Different sizes of the blocks have been considered, ranging from $3$ $\times$ $3$ to 
$8$ $\times$ $8$, as shown in Figure $\ref{Blocks}$. For further details, see the numerical experiments.\\
At last, several different numerical parameters related to the ABCDE analysis of the skin lesions have also been computed, in order to compare the performances of the PBNs with some
standard features usually adopted in classification-type algorithms. To this aim, the suitable for comparisons mask of a skin lesion image has been used to calculate the
principal axes of inertia of the lesion, its orientation, the smallest rectangle containing the mask (with the same orientation) and its inscribed ellipse. These pieces of information
have been used to compute the following parameters.
\begin{itemize}
 \item[(A)] Asymmetry parameters: 
 \begin{itemize}
  \item $Sym1$: the percentage of pixels of the mask whose symmetric with respect to the major axis of inertia lies in the mask;
  \item $Sym2$: the percentage of pixels of the mask whose symmetric with respect to the line perpendicular to the major axis of inertia passing through the barycenter lies in the
  mask;
  \item $Sym3$: the percentage of pixels of the mask  whose symmetric with respect to the barycenter lies in the mask;
  \item $Sym:=(Sym1+Sym2+Sym3)/3$.
 \end{itemize} 
 \item[(B)] Parameters describing properties of the border:
 \begin{itemize}
  \item The form factor of the mask, defined as $4 \pi A /P^2$, where $A$ is the number of pixels of the mask and $P$ is the number of pixels of the boundary;
  \item Haralick's Circularity ($CH$), a standard measure of the compactness of a digital space \cite{Haralick74};
  \item Ellipticity ($Elt$) defined as $4A/(\pi LW)$, where $L$ (respectively $W$) is length of the longest (respectively shortest) side of the smallest rectangle containing the mask;
  \item Eccentricity ($Ecc$), defined as $L/W$, where $L$ and $W$ are as above.
  \end{itemize}
\item[(C)] The Colour Histogram ($Histo$) of the segmented lesion, based on $64$ different colours. 
A Colour Histogram is a vector $\mathbf{H} \in \mathbb{R}^{64}$ whose entries $H_j$ $\in$ $[0,100]$ contain the percentages of pixels of the mask for which the $j$-th colour is
the closest (in the Euclidean metric sense) to their actual $RGB$ values. 
\item[(D)] The diameter ($Diam$) of the lesion, computed as the maximal distance between two pixels of the boundary. Note that using this feature to compare skin lesions requires the
magnification and resolution to be fixed for all the analyzed images.
\item[(E)] The colour Entropy of the skin lesion ($Entr$), a measure of the colour variations among the four regions in which the axes of inertia divide the segmented image 
(\textit{inertial regions}). This is the only non-standard parameter considered in the ABCDE-analysis and is computed as follows: let $\mathbf{H}_0$,...,$\mathbf{H}_4$ be the Colour Histograms of a segmented skin 
lesion related respectively to the global mask and to its four inertial regions. For every $i=0,...,4$ let $H_{ij}$ be the entries of the $i$-th Colour Histogram. Then the
Colour Entropy of the segmented skin lesion is defined as the quantity
$$Entr:=\sum_{i_2>i_1>0} \text{ } \sum_{H_{0j} > 0.05} \frac{|H_{i_2j}-H_{i_1j}|}{H_{0j}}.$$
\end{itemize}

\section{Classification}\label{Classification}
PBNs have a standard structure, the one of superimposed triangles. This has an important outcome, in that the relevant information can be condensed
in the vertices of those triangles, which form the so-called Persistence Diagrams \cite{Fro01}. Comparison of two images (as far as the criterion intrinsic to the filtering
function is concerned) can then be carried out by comparing the sets of these points. Several distances can be defined on the set of PBNs (equivalently, of PDs); a very successful one is the
matching distance (see figure \ref{Bottle} and \cite{D'Amico06} as a general reference). Distance from templates generally produces numbers of
some significance with respect to a classification. Unfortunately, there do not exist archetypal naevi or melanomas, so the task is harder than for classical classification problems.\\
For a fixed database of $N$ images, with $n$ naevi and $m$ mela\-nomas, the software computes $28$ symmetric $N \times N$ matrices, each containing the relative distances between
the images of the database (of course, the main diagonal of these matrices is null):
\begin{itemize}
 \item[\textbullet] $11$ matrices for the $1$-dimensional filtering functions defined in section $\ref{analysis}$ ($1$-dimensional matching distances between the corresponding PBNs);
 \item[\textbullet] $9$ matrices for the $2$-dimensional filtering functions defined in section $\ref{analysis}$ ($2$-di\-mensional matching distances between the corresponding $2$-dimensional
 Size Graphs).
 \item[\textbullet] $8$ matrices for the ABCDE-features of the parameters defined in section $\ref{analysis}$ (of the asymmetry parameters, only the parameter $Sym$, which
 includes the other three, has been used). The relative distance between two images with respect to a generic numerical parameter has been computed as the standard distance in 
 $\mathbb{R}$. A different approach has been used to compute the relative distance between two colour histograms, because a standard Euclidean distance between vectors would
 have not taken into account that some colours are more similar than others. Therefore, the distance between two colour histograms has been computed by minimizing the cost of matching
 the two histograms by means of a minimum weight perfect matching algorithm, where the weight associated to the matching of two colours is the distance of the corresponding colours
 in the $RGB$ space equipped with the Euclidean metric.
\end{itemize}
To perform the image retrieval, all the matrices have been normalized and an average distance has been calculated as a weighted sum of some of these distances. 
The definition and research of an optimal choice for these weights is a challenging problem. Here the following approach has been used: among a very large amount of random attempts,
select the vector of weights that provides the highest diagnostic accuracy with respect to a fixed classifier over the whole database. 
In the numerical experiments, three different classifiers have been considered and analyzed.
\subsection{Classifiers}
Fix a positive integer $k$ and a distance $d$ between the images of the database. For an image $\mathsf{I}$,
compute the first $k$-NN of $\mathsf{I}$ with respect to $d$.\\
Then the first classifier (Standard Classifier) computes the number of melanomas retrieved among the first $k$-NN and 
classifies $\mathsf{I}$ as a naevus if and only if this number is lower than $km/N$. 
The second and the third classifiers (respectively, the Position and the Distance Classifiers) are modified versions of the Standard Classifier that take into account respectively
the positions and the distances of the retrieved images. For example, the Distance Classifier classifies an image as naevus if and only if 
$$\sum_{i=1}^k \frac{\gamma_{N,m,\mathsf{I}_i}}{d(\mathsf{I}, \mathsf{I}_i)} < 0,$$
where $d(\mathsf{I}, \mathsf{I}_i)$ is the distance between $\mathsf{I}$ and the $i$-th data\-base image $\mathsf{I}_i$, and the coefficient $\gamma_{N,m,\mathsf{I}_i}$ is equal to
$(N-m)/N$ if $\mathsf{I}_i$ is a melanoma, and to $-m/N$ otherwise.

\section{Numerical experiments}\label{NumSec}
The image retrieval was tested on well-controlled lesion images. In particular, the experiments were conducted on a dataset containing $107$ atypical melanocytic skin lesions 
undergone to excision and histological examination ($35$ mela\-nomas and $72$ melanocytic naevi).\\
The images were acquired in epiluminescence microscopy with a fixed $16$-fold magnification, with the only selection criterion that the 
lesion had to be entirely visible and with a fixed resolution of $768 \times 576$ pixels.\\
Because of the small dimension of the dataset, it must be emphasized that these experiments cannot lead to significant conclusions, but can nevertheless suggest the road for
future developments of the system.\\
The numerical experiments were performed from four different points of view.

\subsection{Classifiers}
The aim of the first experiment was to compare the results obtained by the classifiers described in section $\ref{Classification}$ on the $28$ features extracted in section $\ref{analysis}$.\\
For the actual computation of the Persistent Betti Numbers on the Global Graphs, the image resolution was reduced by taking averages on squ\-ared blocks of dimensions $4 \times 4$ pixels.
\begin{table}[ht]
\centering
\scriptsize
\begin{tabular}{|c|c|c|c|}\hline
\multicolumn{4}{|c|}{\normalsize \textbf{Test 1: classifiers}}\\ \hline
\textbf{Feature} & \textbf{Std. Class.} & \textbf{Pos. Class.} & \textbf{Dist. Class.}\\ \hline
          & Acc. & Acc.  & Acc.     \\ \hline        
Blue      &76,64 & 80,37 & 76,64    \\
Green     &85,98 & 85,98 & 85,98    \\
Red       &85,05 & 86,92 & 85,05    \\
TV        &75,70 & 74,77 & 74,77    \\
ExcB      &81,31 & 82,24 & 82,24    \\
ExcG      &81,31 & 79,44 & 81,31    \\
ExcR      &82,24 & 80,37 & 81,31    \\
R-G       &80,37 & 78,50 & 80,37    \\
R-B       &81,31 & 77,57 & 80,37    \\
G-B       &79,44 & 81,31 & 81,31    \\
Border    &54,21 & 57,94 & 56,07    \\ \hline
Histo     &75,70 & 74,77 & 77,57    \\
FF        &54,21 & 57,94 & 60,75    \\
CH        &54,21 & 54,21 & 48,60    \\
Sym       &53,27 & 48,60 & 49,53    \\
Elt       &51,40 & 49,53 & 46,73    \\
Ecc       &45,79 & 42,99 & 51,40    \\
Diam      &68,22 & 66,36 & 61,68    \\
Entr      &46,73 & 50,47 & 44,86    \\ \hline
(B,G)     &84,11 & 85,05 & 84,11    \\
(B,R)     &85,98 & 86,92 & 85,98    \\
(G,R)     &85,98 & 86,92 & 85,98    \\
(B,R-G)   &85,05 & 85,05 & 85,98    \\
(G,R-B)   &86,92 & 87,85 & 86,92    \\
(R,G-B)   &85,05 & 85,05 & 85,05    \\
(ExcB,R-G)&85,05 & 86,92 & 85,05    \\
(ExcG,R-B)&79,44 & 78,50 & 80,37    \\
(ExcR,G-B)&85,98 & 85,05 & 85,98    \\ \hline
\end{tabular}
\caption{Numerical results for Test $1$. Accuracy percentage results for $28$ different features.}
\label{RisultatiTest1}
\end{table}
Table $\ref{RisultatiTest1}$ summarizes the results, showing the Accuracy percentages obtained by the single features separately. From the table, it
appears evident that the PBNs obtain better results than the ABCDE features. Moreover, the best performances are usually achieved by the $2$-dimensional PBNs.\\
The classifiers described in section $\ref{analysis}$ have usually similar performances. However, in some particular cases, the differences in the accuracy percentages obtained by the three
classifiers are not negligible. This highlights an important technical difference between image retrieval and classification: whereas classification calculates an automatic possible
diagnosis, image retrieval provides a reliable technical support and remains open to different possible diagnostic results, without replacing the clinician in the classification 
decision.

\subsection{Resolution}
\begin{figure}[t]
 \centering
\includegraphics[width=228pt,height=156pt]{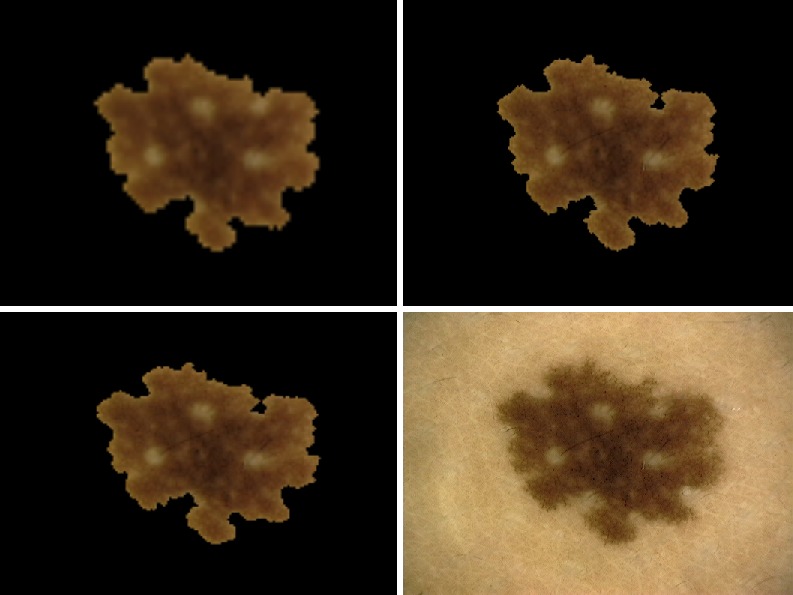}
 \caption{The pictures show the detail deterioration of the image quality caused by the reduction of their resolution. The bottom right picture shows a pigmented
 skin lesion at its original resolution; the other pictures show its segmented version reduced by splitting it into small squared blocks and by taking averages of the three color
 channels on these blocks. Results are displayed when the size of the blocks are $3$ $\times$ $3$ (top right), $4$ $\times$ $4$ (bottom left), $8$ $\times$ $8$ (top left).}
 \label{Blocks}
\end{figure}

The second experiment aimed to compare the results obtained by the Standard Classifier on the features based on Persistent Homology with three different resolutions for computing 
the PBNs defined on the Global Graph.\\
\begin{table}[ht]
\centering
\scriptsize
\begin{tabular}{|c|c|c|c|c|c|}\hline
\multicolumn{6}{|c|}{\normalsize \textbf{Test 2: resolution}}\\ \hline
\textbf{Feature} & \multicolumn{1}{|c|}{$\mathbf{8 \times 8}$} & \multicolumn{1}{|c|}{$\mathbf{4\times 4}$ } & $\mathbf{8}$ \textbf{vs.} $\mathbf{4}$ & \multicolumn{1}{|c|}{$\mathbf{3 \times 3}$} & $\mathbf{4}$ \textbf{vs.} $\mathbf{3}$\\ \hline
          & Acc. & Acc.  &Acc. Diff.     & Acc.  & Acc.Diff. \\ \hline        
Blue      &78,50 &76,64  &\textbf{-1,87} & 78,50 & \textbf{+1,87}\\
Green     &84,11 &85,98  &\textbf{+1,87} & 84,11 & \textbf{-1,87}\\
Red       &85,98 &85,04  &\textbf{-0,93} &87,85  & \textbf{+2,80}\\
TV        &60,75 &75,70  &\textbf{+14,95}&71,03  &\textbf{-4,67}\\
ExcB      &63,55 &81,31  &\textbf{+17,76}&81,31  &\textbf{+0,00}\\
ExcG      &72,90 &81,31  &\textbf{+8,41} &78,50  &\textbf{-2,80}\\
ExcR      &82,24 &82,24  &\textbf{+0,00} &84,11 &\textbf{+1,87}\\
R-G       &83,18 &80,37  &\textbf{-2,80} &85,98 &\textbf{+5,61}\\
R-B       &80,37 &81,31  &\textbf{+0,93} &82,24 &\textbf{+0,93}\\
G-B       &77,57 &79,44  &\textbf{+1,87} &82,24 &\textbf{+2,80}\\ \hline
(B,G)     &84,11 &84,11  &\textbf{+0,00} &n.c.  &n.c.\\  
(B,R)     &83,18 &85,98  &\textbf{+2,80} &n.c.  &n.c.\\
(G,R)     &86,92 &85,98  &\textbf{-0,93} &n.c.  &n.c.\\
(B,R-G)   &85,05 &85,05  &\textbf{+0,00} &n.c.  &n.c.\\
(G,R-B)   &88,79 &86,92  &\textbf{-1,87} &n.c.  &n.c.\\
(R,G-B)   &85,05 &85,05  &\textbf{+0,00} &n.c.  &n.c.\\
(ExcB,R-G)&83,18 &85,05  &\textbf{+1,87} &n.c.  &n.c.\\
(ExcG,R-B)&80,37 &79,44  &\textbf{-0,93} &n.c.  &n.c.\\
(ExcR,G-B)&85,98 &85,98  &\textbf{+0,00} &n.c.  &n.c.\\ \hline
\end{tabular}
\caption{Numerical results for Test $2$. Accuracy percentage results for the PBNs defined on the Global Graphs with different resolutions.}
\label{RisultatiTest2}
\end{table}
Table $\ref{RisultatiTest2}$ shows the Accuracy results obtained with three different block dimensions for the different Size Graphs. Significant oscillations in the accuracy 
results can be noted in many cases. Since they are strongly dependent on the function, a general conclusion cannot be derived. However, it is possible to note that the results 
tend to improve when passing from $8$-dimensional to $4$-dimensional blocks: a more detailed analysis seems to provide better classification results. 
It has to be emphasized that taking averages on $8 \times 8$ blocks produces very low-resolution images, causing an information loss which appears evident
just by displaying the resulting images, as shown in Figure $\ref{Blocks}$. On the other hand, when
reducing the size of the blocks from $4$ to $3$, deteriorations and improvements in the accuracy results are more balanced, suggesting that, at least in this case, too much detail
is not necessary.
Moreover, taking the averages on $3$ $\times$ $3$ blocks produces very large Global Graphs, requiring a huge computational cost. 
Thus, the $2$-dimensional distances between $2$-dimensional Size Graphs were not computed in this case, due to the demanding computational effort. \\
For the sake of completeness, it is worth stating that in this test the behavior of the sensitivity and specificity results was very similar to that of the 
diagnostic accuracy presented here.

\subsection{Number of Nearest Neighbors}
The third experiment was addressed to test the stability of the results with respect to the number of the nearest neighbors.\\
 \begin{figure}[t]
 \centering
 \includegraphics[width=228pt,height=186pt]{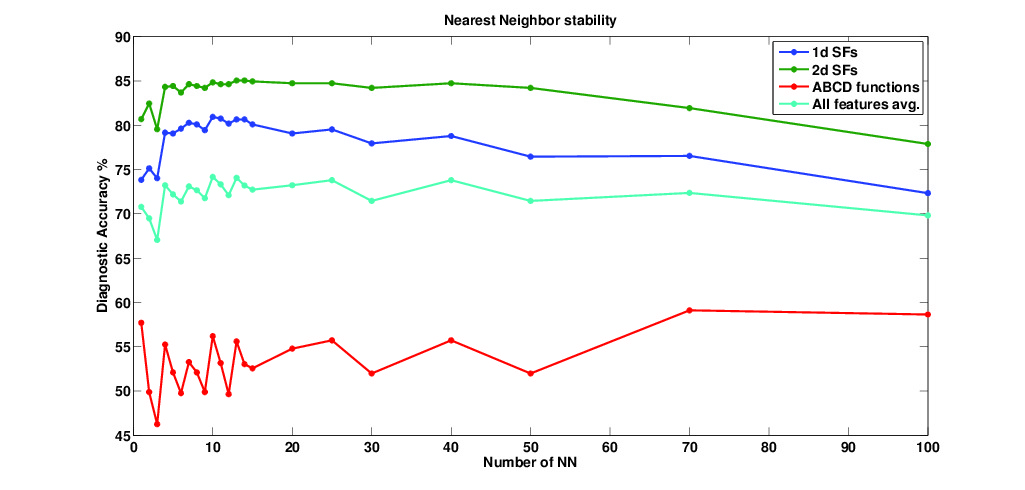}
 \caption{Average Accuracy results obtained by the $28$ functions with respect to the NN number.}
 \label{RisultatiTest3}
\end{figure}
Figure $\ref{RisultatiTest3}$ shows the average behaviors of the diagnostic accuracy with respect to the number of the nearest neighbors. In the test, the Standard Classifier was fixed 
and blocks of size $4 \times 4$ were chosen to simplify the Size Graphs; the figure displays the average results obtained by the $1$-dimensional (blue line) and $2$-dimensional (green line) Persistent Betti Numbers, 
by the ABCDE functions (red line) and by all $28$ features together (cyan line). Although with some differences from case to case, in general the dependency of the results from the
number of NN is not particularly strong. Especially in the case of PBNs, there is a large enough interval of possible choices where the classification results can be considered
equivalent. More precisely, the results do not change significantly if the number of nearest neighbors varies from $6$ to $15$, whereas lower results are achieved when the nearest
neighbors number is too low or too high. A different behavior is observed in the case of the ABCDE functions, but in this case the classification results are too low to drive to any
conclusion.

\subsection{Global distances}\label{GlobDist}
In the last numerical experiment the extracted features were matched together to evaluate the performances of the content based image retrieval system.\\
The Standard Classifier, $10$ nearest neighbors and blocks of size $4 \times 4$ to simplify the Size Graphs were fixed a priori.\\ 
Then three different retrieval systems were implemented as described in section $\ref{Classification}$, by optimizing respectively the $1$-dimensional fun\-ctions, the $2$-dimensional fun\-ctions plus the Border function, and the ABCDE features separately.\\
\begin{table}[ht]
\centering
\small
\begin{tabular}{|c|c|c|c|}\hline
\multicolumn{4}{|c|}{\normalsize \textbf{Test 4: Global Retrieval results}}\\ \hline
Features  & Acc  & Sens  & Spec  \\ \hline
1d PBNs    & 94,39 & 94,29 & 94,44 \\ 
2d PBNs    & 92,52 & 88,57 & 94,44 \\
ABCDE      & 86,92 & 82,86 & 88,89 \\ \hline
\end{tabular}
\caption{Numerical results for Test $4$. Global Retrieval results obtained by matching some of the features together.}
\label{RisultatiTest4}
\end{table}
The classification results, summarized in table $\ref{RisultatiTest4}$, show that the classification systems based on the PBNs were able to distinguish naevi and melanomas 
in a very precise way. In particular, the results obtained by the $1$-dimensional PBNs functions show that only a few of the $107$ images were classified in a wrong way by the 
retrieval based on these functions.\\
Moreover, although the $2$-dimensional PBNs obtained separately the best performan\-ces, they failed to achieve better results
than the $1$-dimensional PBNs in the global case. It is worth mentioning that other numerical experiments, performed with different resolutions and classifiers, led to similar results.\\
On the other hand, despite the very weak results obtained by the ABCDE functions separately, their optimization improved the performances significantly.\\
This remark may be the starting point to consider new possibilities, in which the ABCDE features are used at an early stage to discriminate between very different images and then the $1$-dimensional
PBNs come into play. The authors believe that such hybrid choices should spare computational time without significant losses in the performances.

\subsection{Visual results and discussion}
Since the final goal of the system is the visualization of the image retrieval, it is worth showing some of the results obtained by the retrieval of the $1$-dimensional PBNs functions
described in section $\ref{GlobDist}$.
\begin{figure}[ht]
 \centering
\includegraphics[width=0.7\columnwidth]{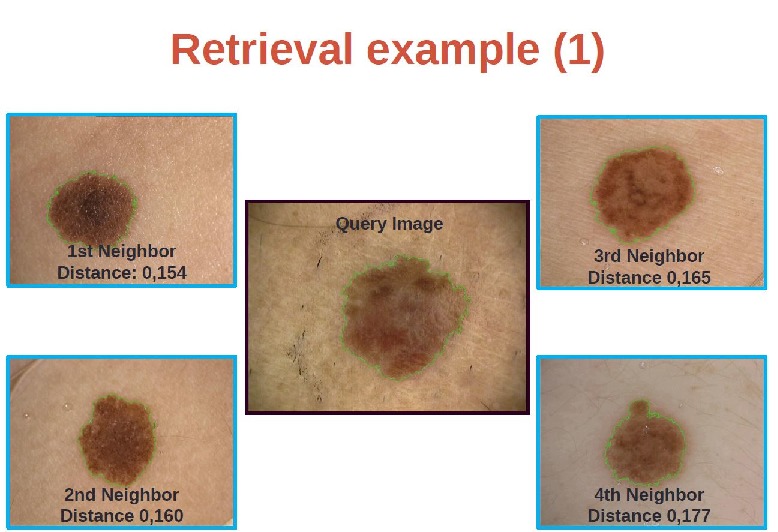}
 \caption{Retrieval of a melanoma with a neighborhood of naevi.}
 \label{Retrieval1}
\end{figure}
Figure $\ref{Retrieval1}$ shows the first four neighbors computed for a query image, histologically diagnosed as melanoma. This is one of the rare cases in which a query image has been
classified by the software in a wrong way. However, a visual inspection reveals how much this diagnosis could be difficult even for an expert clinician: the query image is symmetric,
with a regular border, homogeneously pigmented, and its diameter is smaller than $6$ millimeters. 
The visual retrieval produces a set of neighbors with many clinical features similar to the query, but in this case an automatic classification would be misleading the diagnostic
process.
\begin{figure}[ht]
 \centering
\includegraphics[width=0.7\columnwidth]{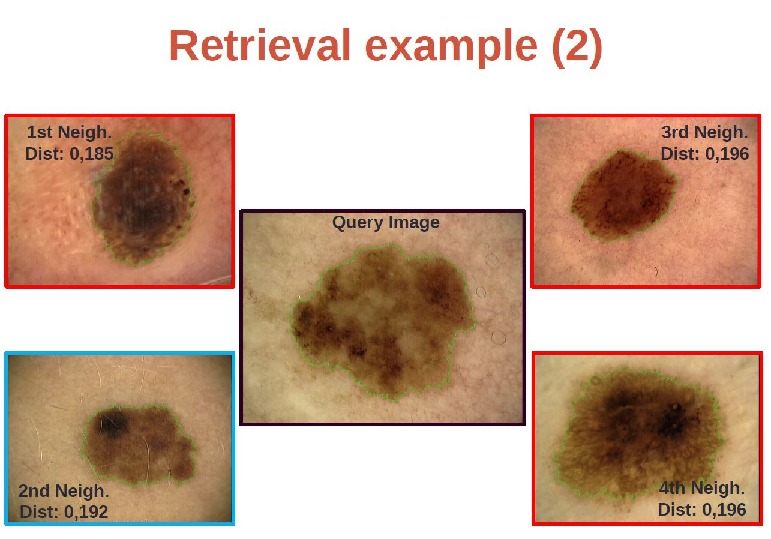}
 \caption{Retrieval of a melanoma with a neighborhood of naevi and melanomas.}
 \label{Retrieval2}
\end{figure}
Figure $\ref{Retrieval2}$ shows the first four neighbors of another melanoma. In this case, the automatic classification is correct, because of the many melanomas present in the
neighborhood of the query. Again, the visual performance is satisfying from the clinical point of view, in the sense that the neighbors share many clinical features with the query,
especially if one takes into account that melanomas tends to be different from each other. Moreover, despite
being a naevus, the second neighbor (bottom left) is similar to the query and could indeed seem a melanoma itself, because of its different colors. An automatic risk computation
could produce a medium-high risk score for this lesion, but the retrieval would justify an even greater concern.\\
These examples show that the retrieval method could prove to be a more useful diagnostic support for the clinician than classification and risk assessment-type algorithms,
suggesting investigating deeper this method and in particular the algorithm proposed in this article.
Summing up, the numerical experiments conducted on this database suggest that the method deserves further investigation. Moreover:
\begin{itemize}
 \item the system appears to be resilient to the reduction of the images' resolution, at least as far as too much detail is not lost. 
 \item The numerical results of the $2$-dimensional Persistent Betti Numbers may not be worth their computational costs for the global distance computation.
 \item Small differences in the choice of the NN number computed by the retrieval system do not seem to affect the accuracy results in a significant way.
 \item Retrieval could prove to be a more efficient diagnostic support than classification.
\end{itemize}

\section{Conclusions}\label{Conclusions}
As pointed out in \cite{Masood13}, at this time, there are no computers that can replace an experienced dermatologist's intuition: a blind trust of the judgement of any computerized 
system could lead the clinician to a failure in the diagnosis. On the other hand, computerized systems can prove to be an
important diagnostic support for the dermatologist in the diagnostic phase.\\
Starting from this consideration, in this paper the authors propose a new $k$-NN algorithm for the image retrieval of skin lesions. The main novelty of the algorithm is the extensive 
use of Persistent Homology theory, with new tools with respect to the previous works \cite{D'Amico04,FerriStanga2010,Stanga05}, which enable a global analysis of the skin lesion.
These tools have been tested on a dataset of $107$ melanocytic lesions, with promising results, leading to the conclusion that this method deserves further investigation and tests.

\section*{Acknowledgements}
Support by ARCES, CA-MI S.r.l., IRST-IRCCS and the National Institute of High Mathematics ``F. Severi'' (INdAM) is gratefully acknowledged.

\bibliographystyle{spmpsci}      
\bibliography{mybibfile} 

\end{document}